 \documentclass[12pt,preprint,usenatbib]{aastex}  

\def\dsm{$\mathrm{M}_\odot$}

\shorttitle{CMDs of NGC1831, NGC1868 and NGC2249} \shortauthors{Li Z.M. el al.}

\begin{document}


\title{FAILURE OF ROTATION AND NOTABLE SUCCESS OF AGE SPREAD TO REPRODUCE THE CMDS OF STAR CLUSTERS NGC 1831, NGC 1868 and NGC 2249}

\author{Zhongmu Li\altaffilmark{1,2}, Caiyan Mao\altaffilmark{1},Liyun Zhang\altaffilmark{3}}

\altaffiltext{1}{Institute for Astronomy and History of Science and
Technology, Dali University, Dali 671003, China;
zhongmu.li@gmail.com} \altaffiltext{2}{Key Laboratory for the
Structure and Evolution of Celestial Objects, Chinese Academy of
Sciences, Kunming 650011, China} \altaffiltext{3}{College of Science/Department of Physics \& NAOC-GZU-Sponsored Center for Astronomy Research, Guizhou University, Guiyang 550025, China}

\begin{abstract}
We study the high-quality CMDs of three star clusters, NGC 1831, NGC 1868 and NGC 2249 in detail,
via the two most likely causes (stellar rotation and age spread) for CMDs with extended main-sequence turn-offs.
The results show evident failure of stellar rotation (including resolved and unresolved binary stars) to interpret the CMDs of three star clusters,
and the unexpected success of age spread.
In particular, the special structures of turn-off and red clump parts cannot be generated by stellar rotation,
but age spread perfectly reproduces the observed features. This suggests that these three clusters contain multiple populations of stars,
rather than a single population of rotating stars and binaries. The thick subgiant branch of NGC 1831 gives the strongest support to this.
The results demonstrate that stellar rotation cannot save the widely accepted view (simple population) of star clusters,
and an extended star formation history is needed for explaining the observed CMDs.
In addition, this work shows that a narrow subgiant branch does not correspond to a simple population.
The judgement of simple population in NGC 1651 by a previous work is not necessarily reliable.
We anticipate our assay to be a starting point for more precise study of Hertzsprung-Russell diagrams of star clusters.
This involves many factors, such as binaries, rotating stars, and star formation history (SFH).
\end{abstract}


\section{Introduction}
Star clusters are always thought of as simple celestial bodies that each consists of a simple population of stars with the same age.
In this picture, all stars of a cluster distribute on a simple curve in the Hertzsprung-Russell diagram (i.e., isochrone).
Surprisingly, however, the observed counterpart of Hertzsprung-Russell diagrams, namely color-magnitude diagrams (CMDs),
of intermediate-age (0.5 to 2 Gyrs) clusters unexpectedly exhibit special shapes, including extended main-sequence turn-offs (eMSTO) and red clumps (eRC),
which are far from simple curves\citep{mack07,mack08, milo09,Piatti13,girardi2009,Girardi13}. This is obviously against to the accepted picture.
Great efforts have been made to explore why this should be: a spread of stellar age and stellar rotation being finally thought as the two most likely candidates\citep{bert03,
piot05,piot07,mack07,derc08,mack08, glat08,mucc08,bast09,goud09,milo09,milo10,rube10,Girardi11,Goudfrooij11a,Goudfrooij11b,kell11,Rubele11,Yang11,Li12,platais12,Girardi13,Richer13,goud2014,Li2014a,Li2014b,Li2015}.
However, the answer is still uncertain, because both candidates are capable of explaining the observations of certain clusters.

A few recent works (e.g., \citealt{Li2014b}) argue stellar rotation as the cause for special CMDs of star clusters,
but there are no systematic comparisons between the rotation models and the observations. \cite{Li2015}
supplies such a work and indicates that rotation is of the ability to form the CMD of NGC 1651.
However, it is impossible to conclude that rotation can lead to all CMDs with eMSTO,
because rotation possibly affects different stellar populations in different ways.
It is therefore necessary to study the CMDs of various star clusters in more detail.
Because \cite{Li2015} have studied a 1.5\,Gyr-old cluster, NGC 1651,
this work attempts to study some other clusters with younger ages.
The advantage of studying younger clusters is that stellar rotation may affect such clusters much more slightly,
because only massive stars reach turn-off region at young ages (see also \citealt{georgy2013,yang2013}).
Three clusters in the Large Magellanic Cloud (LMC), i.e., NGC 1831, NGC 1868 and NGC 2249,
are finally chosen for this work, according to their special CMD
structures and small uncertainties in magnitudes and colors. They are shown to be ideal targets for our purpose.
Compared to previous works, e.g., \cite{Li2014a} and \cite{Correnti2014},
this work has two obvious progresses: considering the observational errors accurately and testing both two potential answers for eMSTO in detail.
The effects of resolved and unresolved binaries are taken into account via a proper way.
A well calculated database \citep{georgy2013} of rotating stars
and some carefully determined distributions of rotation rate of stars \citep{royer07} are taken for considering the effects of stellar rotation.
Surprisingly, the results are shown to disagree with that of NGC1651:
rotation being far from explaining the CMDs of three clusters.

The structure of this paper is as
follows. Sections 2 and 3 introduce the observed and theoretical
CMDs. Section 4 introduces the technique of CMD fitting shortly.
Section 5 presents the best-fit results from two models.
Finally, Section 6 draws a brief summary.

\section{Observed CMDs}
\subsection{Data and Photometry}
The data set of NGC 1831, NGC 1868 and NGC 2249 are retrieved from the $Hubble$ $Space$ $Telescope$ ($HST$)
archive, which were observed using the Wide Field Planetary Camera 2 (WFPC2).
The exposures of NGC1831 and NGC 1868 are 800 and 900 seconds in both F555W
and F814W ($V$ and $I$) filters. The images of NGC 2249 are observed in filters F439W and F555W ($B$ and $V$),
with exposures of 230 and 120 seconds respectively.
The data are handled using a stellar photometry package (HSTphot, \citealt{Dolphin2000})
specially designed for use with $HST$/WFPC2 images.
In order to reduce the contamination of field stars,
only some luminous central parts are taken for our study.
Because the surface brightness of a cluster usually decreases with increasing distance from the center \citep{mack03},
we find the cluster center by searching for the part (with a radius of 0.4\,arcsec) with the largest surface brightness.
The surface brightness of the most luminous part is denoted by $f_{\rm c}$.
Then the central part with surface brightness more luminous than 0.3 $f_{\rm c}$ is taken for each cluster.
The results seem very good and suitable for our work, as shown in panels $c$, $f$ and $i$ of Fig. 2.

\subsection{Corrections for Incompleteness and Field Contamination}
Star incompleteness is mainly caused by crowding. We therefore perform a series of artificial
star tests (ASTs) to characterize the completeness of stars (see also \citealt{rube10}).
The results are shown in Fig. 1. As we see, the CMDs of three clusters are of high completeness,
as most CMD parts have completeness greater than 0.8.
According to the star incompleteness, some random stars (i.e., compensated stars, as shown in panels $a$, $d$ and $g$ of Fig. 2),
are added into the observed CMDs to correct for star incompleteness.

Besides this, a correction for field star contamination is made.
Some areas that are distant from the cluster center and with surface brightness less than 0.1~$f_{\rm c}$ are defined as field regions.
Some field stars in the central part are statistically picked out from the original CMDs following an assumption that field stars distribute uniformly in the whole image.
This method is relatively reliable, because the data of cluster and field are taken from the same image.
A similar method was also used by \cite{Li2014b}.
Although it is possible that very few cluster stars are included in the chosen field part, the number is shown to be very small,
as only a few stars are removed (refer to panels $b$, $e$ and $h$ of Fig. 2).
Thus the treatment of correction for field contamination does not affect the final results.

\begin{figure} 
\centering
\includegraphics[angle=-90,width=0.9\textwidth]{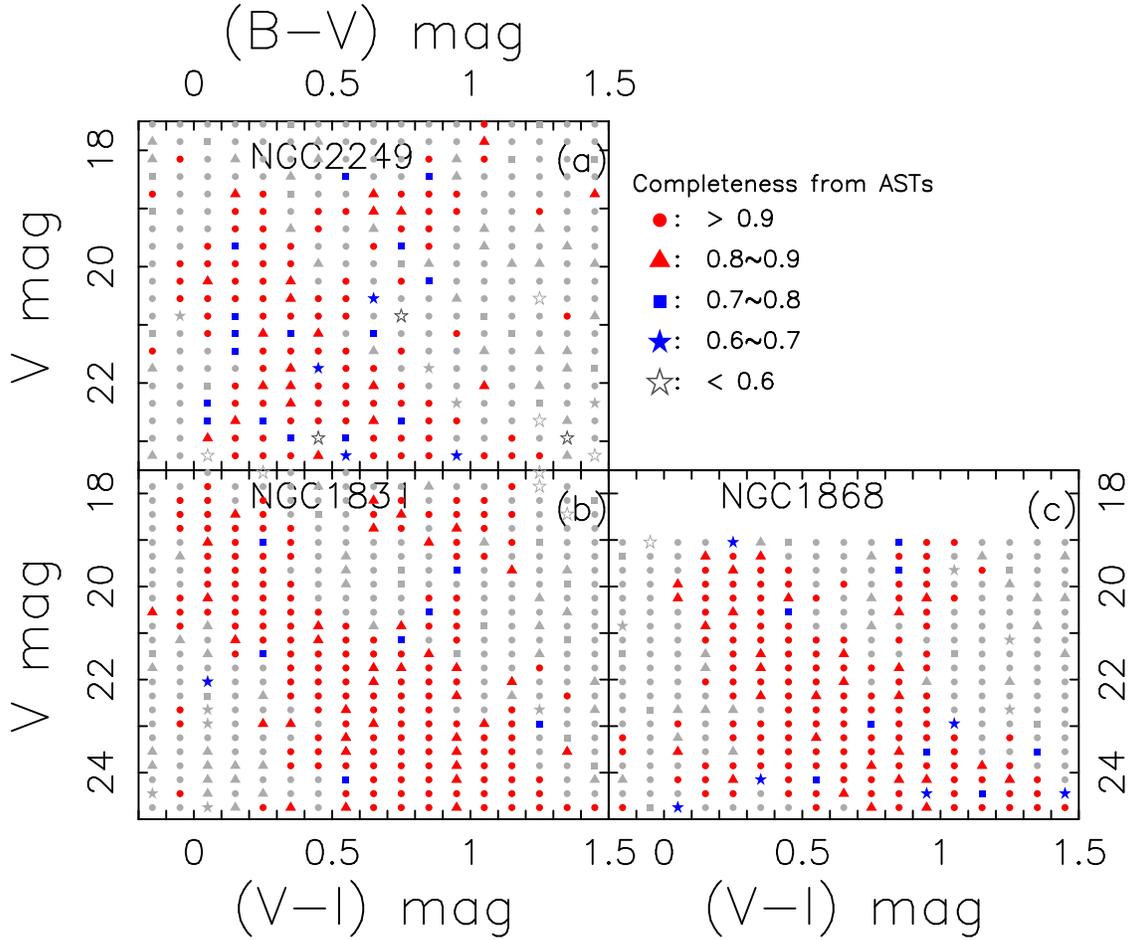}
\caption{Star completeness from ASTs.  The results for where observed stars are located are highlighted by red, blue and black colors,
and that for the other part is plotted in light grey.}
\end{figure}

\begin{figure} 
\centering
\includegraphics[angle=-90,width=0.9\textwidth]{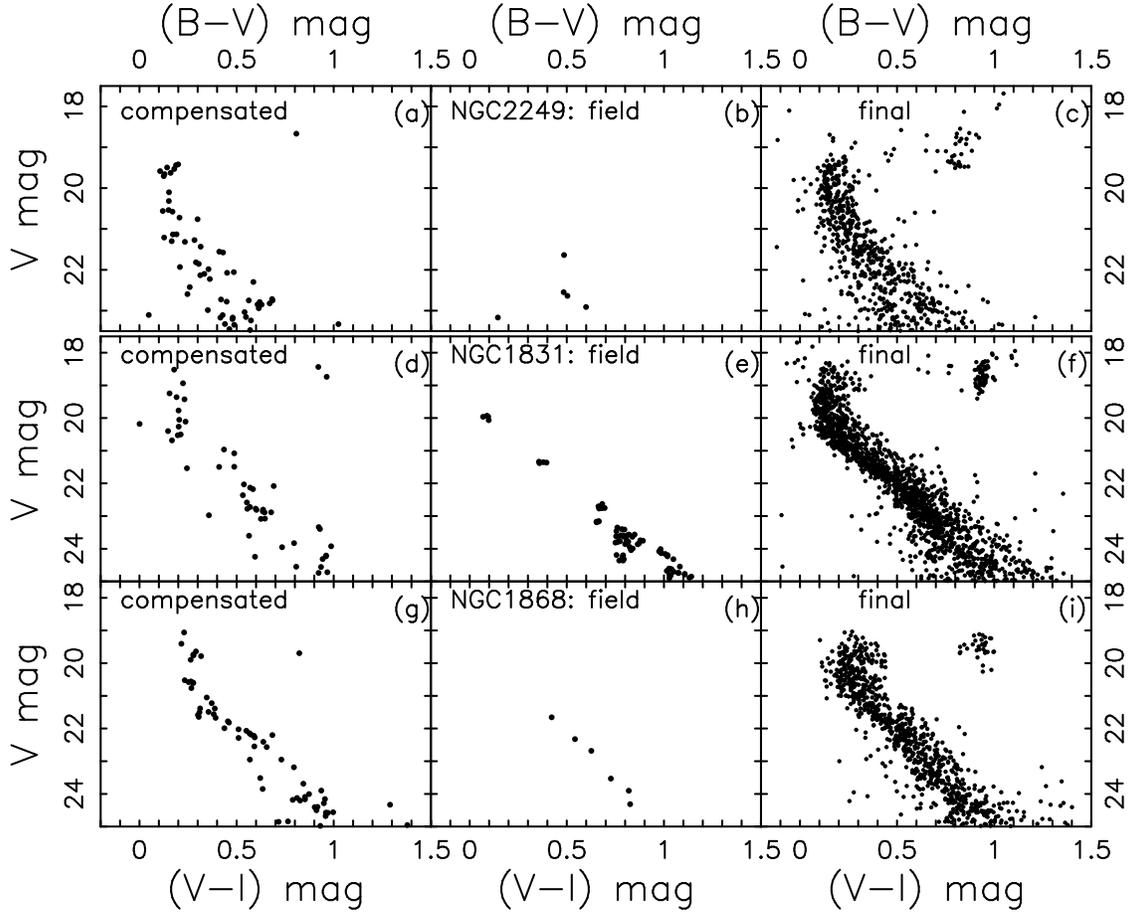}
\caption{Corrections for star incompleteness and field star contamination.
The first column shows compensated stars that are added into the CMDs according to star completeness,
while the second column shows field stars that are removed from the observed CMDs. The third column shows the final CMDs after corrections.
The star numbers of NGC 1831, 1868 and 2249 are 1460, 870 and 652 respectively.}
\end{figure}

\subsection{Photometric Errors}
The photometric errors are mainly caused by crowding and the data handling process.
Such errors can be well estimated by ASTs.
In this work, more than 10$^{\rm 5}$ artificial stars are generated to test the errors of magnitudes and colors.
Fig. 3 shows the main results. We find that the errors in observed magnitudes and colors of stars are very small.
The values for most stars are less than 0.05\,mag. This is good for detailed tests to various models of eMSTO.
Note that the errors reported by ASTs are significantly larger than those given directly by HSTphot (see \citealt{Li2014a} for comparison).

\begin{figure} 
\centering
\includegraphics[angle=-90,width=0.9\textwidth]{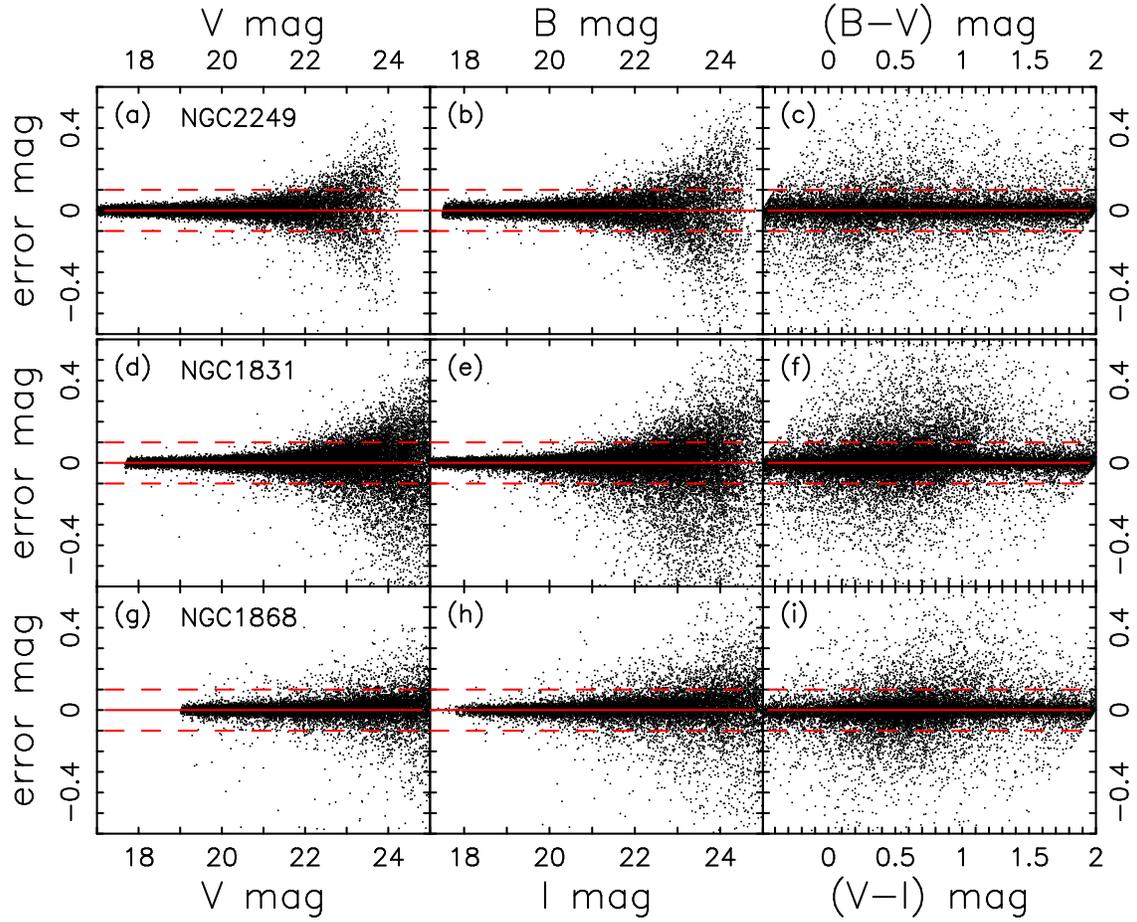}
\caption{Errors in magnitudes and colors. Black points show the errors as functions of magnitudes and colors.
Solid red lines mean no error, while dashed red lines denote an error of 0.1\,mag.}
\end{figure}

\section{Synthetic CMDs}
The construction of synthetic CMDs is similar to some previous works
(\citealt{zhang04,Li2008mn,Li2008apj,Li2012,Li12,Jiang2014,Zhang2015,Li2015}).
In brief, the initial mass function (IMF)
of \cite{Salpeter1955} with lower and upper mass limits of 0.1 and
100 \dsm{} respectively is taken.
Binaries are taken into account, because every cluster possibly contains some binaries (see e.g., \citealt{elson98,Li2015}).
The mass of the primary component of a binary is generated following the selected IMF,
and the mass of the secondary component is calculated by taking a random
secondary-to-primary mass ratio ($q$), which obeys a uniform
distribution within 0--1. The eccentricity ($e$) of each binary is
given randomly within 0-1. The separations ($a$) of two components
are given by:
\begin{equation}
an(a)=\left\{
 \begin{array}{lc}
 \alpha_{\rm sep}(a/a_{\rm 0})^{\rm m} & a\leq a_{\rm 0};\\
\alpha_{\rm sep}, & a_{\rm 0}<a<a_{\rm 1},\\
\end{array}\right.
\end{equation}
where $\alpha_{\rm sep}\approx0.070$, $a_{\rm 0}=10R_{\odot}$,
$a_{\rm 1}=5.75\times 10^{\rm 6}R_{\odot}=0.13{\rm pc}$ and
$m\approx1.2$ \citep{han95}. This leads to about 50\% stars
in binaries with orbital periods less than 100 yr,
but the binary fraction can be changed by removing some random binaries.

The evolution of stars is computed using the rapid stellar evolution code of \cite{Hurley98} and \cite{Hurley02}
(hereafter Hurley code). Most binary interactions such as mass transfer,
mass accretion, common-envelope evolution, collisions, supernova
kicks, angular momentum loss mechanism, and tidal interactions are
taken into account. The typical uncertainty of this code is about 5
per cent in stellar luminosity, radius and core mass, and it affects the results slightly.

Because Hurley code does not take stellar rotation
into account, the effects of rotation on effective
temperature, luminosity and main-sequence lifetime are taken into account for massive ($>$ 1.7
\dsm{}) stars, additionally. A recent database of evolution of rotating stars \citep{georgy2013} is adopted.
The database of \cite{georgy2013} is particularly
useful for constructing synthetic populations of stars, accounting
for mass, rotation, and metallicity distributions. It includes an accurate computation
of the angular-momentum and stellar-wind anisotropy and therefore is relatively reliable.
The changes of surface temperature and luminosity, which are caused by stellar rotation, are calculated by
comparing the evolutionary tracks of rotating stars to those of non-rotating ones.
Because of centrifugal forces, rotating stars behave like non-rotating stars of lower mass (see also \citealt{ekstrom2008,meynet2000}).
We therefore calculate the main-sequence lifetime change of rotating stars using the results of \cite{georgy2013}.
Because of the lack of evolutionary tracks of stars with $Z$ = 0.008, stellar metallicity is interpolated to
the typical value of LMC clusters.

Some random values are assigned to the rotation rates of members of a population of rotating stars,
following the observed results of \citet{royer07} (similarly, \citealt{zorec2012}),
because there are obvious distributions for the rotation rates of stars. Finally, the stellar evolutionary parameters ([Fe/H], T$_{eff}$, $\log g$, $\log L$)
are transformed into colors and magnitudes using the well-calibrated atmosphere library of \cite{leje98}.
The photometric errors derived from ASTs are applied to the simulated stars when comparing synthetic CMDs with the observed ones.
This enables doing real comparisons between synthetic and observed CMDs.

\section{CMD Fitting Technique}
Following a previous work \citep{Li2015}, we use a new method to fit the CMDs.
The new technique searches for the best-fit parameters by
comparing the star fraction in every part (or grid) of a CMD.
An observed CMD is divided into 1500 grids via selected color and magnitude intervals.
The goodness of fit is judged by the average difference of grids that contain observed or theoretical stars (hereafter effective grids),
when taking the weight of each grid into account.
A parameter, i.e., weighted average difference ($WAD$), is defined to denote the goodness of fit.
$WAD$ can be calculated by
\begin{equation}\label{eq3}
    WAD = \frac{\Sigma [{\omega_i.|f_{ob}-f_{th}|}]}{\sum\omega_i}~,
\end{equation}
where $\omega_i$ is the weight of $i$th grid, and
$f_{ob}$ and $f_{th}$ are star fractions of observed and theoretical
CMDs in the same grid. $\omega_i$ is calculated as
\begin{equation}\label{eq4}
    \omega_i = \frac{1}{|1-C_i|} = \frac{1}{\sigma_i}~,
\end{equation}
where $C_i$ is the completeness of $i$th grid, and $\sigma_i$ means star fraction uncertainty.
As shown by \cite{Li2015}, WAD fitting is able to find the best-fit parameters accurately, and this technique is as reliable as $\chi^2$ fitting.
Because there is no evidence for significant spread of metallicity of stars in a cluster,
we fix stellar metallicity to the typical value of LMC clusters, $Z$ = 0.008.
The effects of resolved and unresolved stars have been taken into account according to the separations of binary components and the discrimination of $HST$.
Our new code, ``Binary Star to Fit for CMD'' ($BS2fit for CMD$) \citep{Li2015}, is used for CMD analysis.

\section{Results}
The best-fit parameters of three star clusters are listed in Table 1.
As we see, different results are obtained from simple (SSP) and composite (CSP) stellar populations.
The stellar age of a cluster can be different as large as 0.5\,Gyr (for NGC 1868) when taking different models.
Although two kinds of models measure similar distances for NGC 1831,
they report different values for NGC 1868 and NGC 2249.
Some of these values are similar to while some others are different to previous findings (e.g., \citealt{Li2014a} and \citealt{Correnti2014}).
Meanwhile, SSP and CSP models show various color excesses for all the three clusters.
However, similar binary (those with orbital period less than 100\,yr, rather than interactive binaries) fractions are found by various models.

Figs. 4 and 5 show the comparisons of best-fit and the observed CMDs of three star clusters.
Here CMDs with (Fig. 4) and without (Fig. 5) observational errors are presented respectively to make the comparisons clearer.
It is obvious that SSP models consisting of rotating stars with a single age cannot fit the CMDs of the three clusters well.
It is natural for NGC 1831, because only stars more massive than 2.58 \dsm{} reach turn-off (around 0.5\,Gyr),
but rotation affects such massive stars only very slightly compared to less massive ones,
on account of their large gravities \citep{georgy2013,yang2013}.
The presence of eMSTO of NGC1831 strongly suggests that rotation may not be the main reason of eMSTO.
In addition, although stellar rotation extends some the turn-offs of NGC 2249 and NGC 1868,
it is far from the observation.
We found that the change of logarithmic luminosity caused by full rotation (rotation rate is 1)
is only one third of that caused by a 100 million-year spread of stellar age, for stars less massive than 2.5 \dsm{}.
Furthermore, if a cluster consists of rotating stars with the same age,
it is difficult to imagine how separate turn-offs of NGC 2249 (see also another work, \citealt{Correnti2014})
and the bizarre turn-off structure of NGC 1868 form,
because most stars rotate slower than 250 km s$^{-1}$ (rotation rate is less than about 0.8)
and the rotation rate distributes continuously, rather than discretely (for velocity $>$ 100 km s$^{-1}$) \citep{royer07,zorec2012}.

Meanwhile, the CSP models with multiple stellar populations fit to the observations perfectly.
Most CMD details, including extended main-sequence, turn-off (TO), and peculiar red clump (RC),
are reproduced much better compared to SSP models. In particular, the strange TO structures of NGC 2249 and NGC 1868 are reproduced, naturally.
It seems impossible to be an accident. This implies that the three clusters contain multiple populations of stars.
We find other strong evidence from NGC 1831.
As we see in Figs. 4 and 5, the comparison of the distribution of subgiant-branch stars (pentagrams) to certain isochrones with various ages reports a
minimum age spread of 300\,Myrs. Because the average magnitude and color errors for stars brighter than 20\,mag are only 0.002 and 0.003\,mag
(less than 0.5\% of the observed subgiant-branch width),
and similar subgiant-branch stars are also shown by another work \citep{Li2014a},
this evidence strongly excludes the simple population explanation of NGC 1831.
Note that every subgiant-branch star is valuable, as such stars are difficult be observed.

We found that the subgiant-branch stars of some populations cannot be observed
if these population contain only a small number (e.g., a few hundreds) stars,
as stars evolve so rapidly (2.8--12\,Myr for the three clusters here) at subgiant stage.
Therefore, subgiant-branch stars cannot give a reliable constraint on the maximum age spread of a cluster.
The CMD of NGC 1868 convincingly confirms this, as there are no obvious subgiant-branch
stars in either the observed or best-fit CMDs, even though the theoretical population contains 6 star bursts.
One can check into the blue box region of Figs. 4 and 5, where a star burst means an SSP with a fixed age.
In this case, the conclusion of a previous work \citep{Li2014b} is not necessarily reliable.
The narrow subgiant branch of clusters like NGC 1651 does not mean there is no significant age spread in these clusters.
One of our works \citep{Li2015} has shown that even though NGC1651 contains 5-6 different populations of stars,
the subgiant branch can be as narrow as a population with a single age.

\begin{table}
 \caption{Best-fit parameters respectively from various models based on stellar rotation (SSP) and age spread (CSP) assumptions. All models have a metallicity of $Z$ = 0.008.
SSP refers to the simple population of rotating stars with the same age, while CSP is the composite population of non-rotating stars with various ages.
$(m-M)_0$ and $E(V-I)$ are distance modulus and color excess in mag.
$f_{\rm bin}$ and $f_{\rm rot}$ are fractions of binaries and rotators.
SFH lists the ages (in Gyrs) of component populations and the mass fraction of each component population in the cluster.
The less the goodness indicator (WAD), the better the fit to an observed CMD. $\chi^{2}$ fitting gives the same results.}
 \label{symbols}
 \begin{tabular}{l}
  \hline
 SSP fit of NGC 1831 (WAD=0.00231):\\
$(m-M)_0$=18.39,  $E(V-I)$=0.08, $f_{\rm bin}$=0.6,$f_{\rm rot}$=1.0\\
SFH: 0.6(100\%)\\
 CSP fit of NGC 1831 (WAD=0.00197):\\
$(m-M)_0$=18.39,  $E(V-I)$=0.10, $f_{\rm bin}$=0.6,$f_{\rm rot}$=0\\
SFH: 0.4(5\%), 0.5(30\%), 0.6(25\%), 0.7(20\%), 0.8(20\%)\\
  \hline
 SSP fit of NGC 1868 (WAD=0.00438):\\
$(m-M)_0$=18.84,  $E(V-I)$=0.12, $f_{\rm bin}$=0.6,$f_{\rm rot}$=1.0\\
SFH: 0.8(100\%)\\
 CSP fit of NGC 1868 (WAD=0.00391):\\
$(m-M)_0$=18.50,  $E(V-I)$=0.10, $f_{\rm bin}$=0.6,$f_{\rm rot}$=0\\
SFH: 0.8(20\%), 0.9(20\%), 1.0(20\%), 1.1(15\%), 1.2(15\%), 1.3(10\%)\\
  \hline
 SSP fit of NGC 2249 (WAD=0.00497):\\
$(m-M)_0$=18.51,  $E(B-V)$=0.01, $f_{\rm bin}$=0.6,$f_{\rm rot}$=1.0\\
SFH: 1.0(100\%)\\
 CSP fit of NGC 2249 (WAD=0.00445):\\
$(m-M)_0$=18.66,  $E(V-I)$=0.05, $f_{\rm bin}$=0.5,$f_{\rm rot}$=0\\
SFH: 0.8(35\%), 0.9(35\%), 1.1(30\%)\\
  \hline
 \end{tabular}
 \end{table}

 \begin{figure} 
\centering
\includegraphics[angle=-90,width=0.9\textwidth]{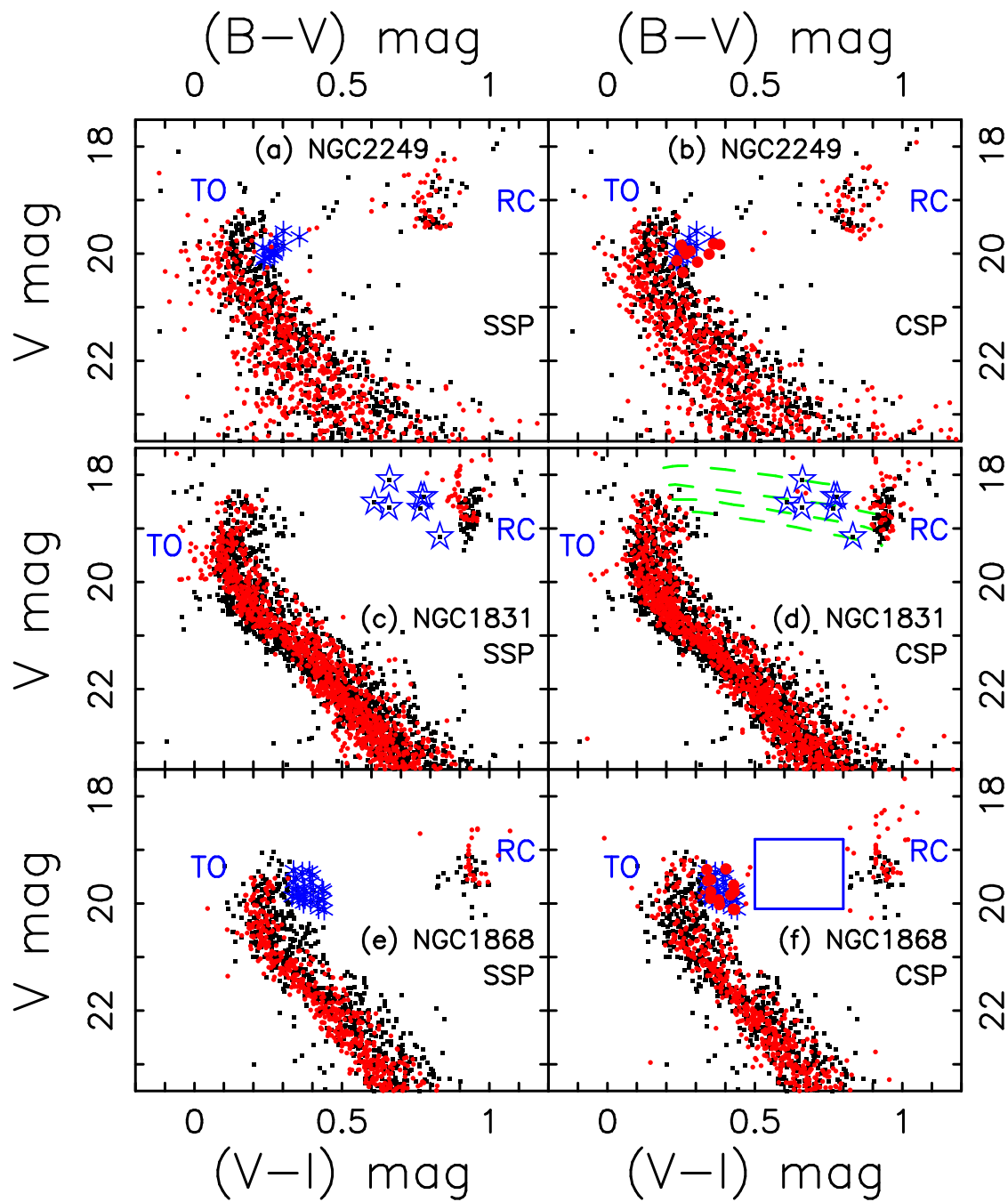}
\caption{Comparison of best-fit (red points) and the observed (black squares) CMDs.
Blue stars and pentagrams highlight some stars used for special comparison.
``TO'' and ``RC'' denote turn-off and red clump parts.
Green dashed lines from top to bottom show isochrones with ages of 0.8, 0.9, 1.0 and 1.1\,Gyrs.
Observational errors from ASTs have been applied to the theoretical CMDs, and
the best-fit and observed CMDs contain similar (difference $<$ 20) objects.}
\end{figure}

\begin{figure} 
\centering
\includegraphics[angle=-90,width=0.9\textwidth]{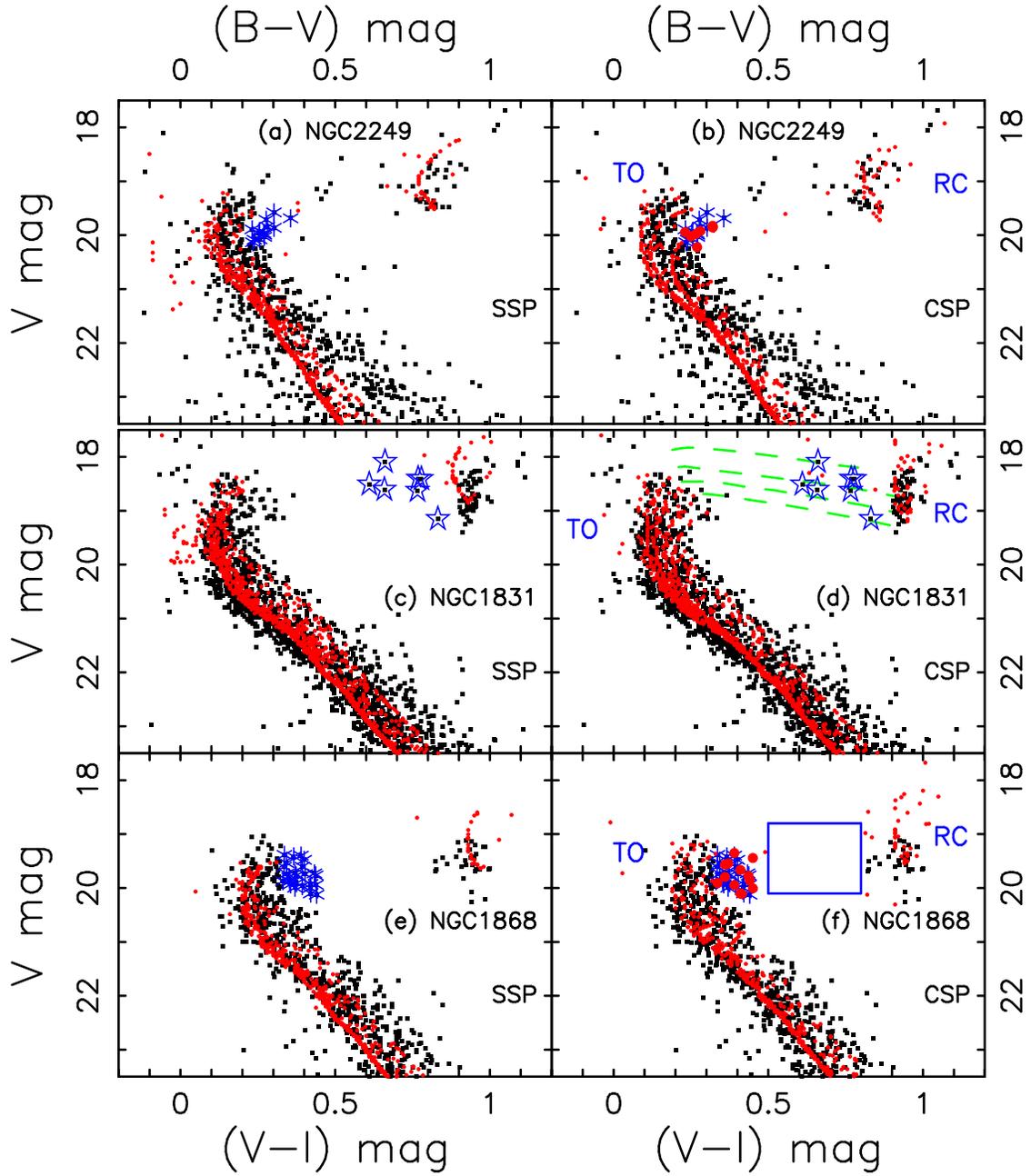}
\caption{Similar to Fig. 4, but observational errors are not applied to the theoretical CMDs.}
\end{figure}

\section{Conclusions}

We investigate the high-quality CMDs of three LMC star clusters, NGC 1831, NGC 1868 and NGC 2249,
using the two most likely causes (stellar rotation and age spread) of eMSTO of intermediate-age clusters.
Obvious failure of simple stellar populations (SSP) to reproduce the observed CMDs,
and the notable success of composite stellar populations (CSP) with multiple components, are shown clearly.
Together with an unquestioned evidence for multiple stellar populations of NGC 1831,
we can conclude that very likely there are multiple populations of stars in clusters NGC 1831, NGC 1868 and NGC 2249.
Therefore, within the two candidates for CMDs with extended turn-offs,
age spread is possibly the only or dominant reason, although stellar rotation may affect some CMDs,
at least for the three clusters investigated by this work.
Even only NGC 1831 is a composite stellar population, the classical image of star clusters,
i.e., the simple stellar population image, needs to be upgraded.
We also argue that a previous result \citep{Li2014b} about the simple population of NGC 1651 is likely to be unreliable.
Our results give support to other works such as \cite{Correnti2014} and \cite{goud2014}.
Although the result poses a major challenge to our understanding of star formation and dynamics of star clusters,
the excellent fits to the observed CMDs are not accident.
It is possible that some important knowledge of star formation and dynamics of star clusters is not discovered yet.
In this case, more detailed and deeper studies about the types of stellar populations and the star formation processes of clusters are needed.

\acknowledgments  This work has been supported by the Chinese National
Science Foundation (Grant No. 11203005), Open Project of Key Laboratory for the Structure and Evolution of Celestial Objects,
Chinese Academy of Sciences (OP201304).


\end{document}